\documentclass[runningheads]{llncs}
\usepackage[T1]{fontenc}
\usepackage{graphicx}
\usepackage{cite}
\usepackage{amsmath}
\usepackage{amsfonts}
\usepackage{bm}%加粗
\usepackage{multirow}
\usepackage{makecell}
\usepackage{booktabs} 
\usepackage{lipsum}
\usepackage{mathrsfs}
\usepackage{caption}
\usepackage{hyperref}
\usepackage{subcaption}
\usepackage{bbding}
\usepackage{orcidlink}
\begin{document}
\title{MMFusion: Multi-modality Diffusion Model for Lymph Node Metastasis Diagnosis in Esophageal Cancer}
\titlerunning{MMFusion}
\author{Chengyu Wu\inst{1,\dagger}\orcidlink{0009-0009-0450-8649} \and
Chengkai Wang\inst{2,\dagger}\orcidlink{0009-0000-2690-9608} \and
Yaqi Wang\inst{3}\orcidlink{0000-0002-4627-3392} \and Huiyu Zhou\inst{4}\orcidlink{0000-0003-1634-9840} \and Yatao Zhang\inst{1}\orcidlink{0000-0002-6152-0806} \and Qifeng Wang\inst{5}\orcidlink{0000-0001-8544-9007} \and Shuai Wang\inst{6,7}\orcidlink{0000-0003-3730-6401}}
% \inst{6,7(}\Envelope\inst{)}
%
\authorrunning{C. Wu et al.}
\institute{Department of Mechanical, Electrical and Information Engineering, Shandong University, Weihai, China. \and
School of Management, Hangzhou Dianzi University, Hangzhou, China. \and
College of Media Engineering, Communication University of Zhejiang, Hangzhou, China. \and School of Computing and Mathematical Sciences, University of Leicester, Leicester, United Kingdom. \and Department of Radiation Oncology, Sichuan Cancer Hospital and Institution, Sichuan Cancer Center, School of Medicine, Radiation Oncology Key Laboratory of Sichuan Province, University of Electronic Science and Technology of China, Chengdu, China. \and School of Cyberspace, Hangzhou Dianzi University, Hangzhou,  China \and Suzhou Research Institute of Shandong University, Suzhou,  China. \\ \email{wuchengyu@mail.sdu.edu.cn, wangck@hdu.edu.cn, wangyaqi@cuz.edu.cn, 
hz143@leicester.ac.uk,
zytboy@sdu.edu.cn, 
wangqifeng@uestc.edu.cn,
shuaiwang.tai@gmail.com}
}

% \institute{Anonymous Organization\\
% \email{**@******.***}}
\renewcommand{\thefootnote}{\fnsymbol{footnote}}
\footnotetext{$\dagger$ These authors contributed equally to this work.}

\maketitle    % typeset the header of the contribution

\begin{abstract}
Esophageal cancer is one of the most common types of cancer worldwide and ranks sixth in cancer-related mortality. Accurate computer-assisted diagnosis of cancer progression can help physicians effectively customize personalized treatment plans. Currently, CT-based cancer diagnosis methods have received much attention for their comprehensive ability to examine patients' conditions. However, multi-modal based methods may likely introduce information redundancy, leading to underperformance. In addition, efficient and effective interactions between multi-modal representations need to be further explored, lacking insightful exploration of prognostic correlation in multi-modality features. In this work, we introduce a multi-modal heterogeneous graph-based conditional feature-guided diffusion model for lymph node metastasis diagnosis based on CT images as well as clinical measurements and radiomics data. To explore the intricate relationships between multi-modal features, we construct a heterogeneous graph. Following this, a conditional feature-guided diffusion approach is applied to eliminate information redundancy. Moreover, we propose a masked relational representation learning strategy, aiming to uncover the latent prognostic correlations and priorities of primary tumor and lymph node image representations. Various experimental results validate the effectiveness of our proposed method. The code is available at \hyperlink{https://github.com/wuchengyu123/MMFusion}{https://github.com/wuchengyu123/MMFusion}.

\end{abstract}
\section{Introduction}
Esophageal cancer, a leading cause of cancer-related mortality globally, presents significant challenges in diagnosis and treatment planning~\cite{sung2021global}. Physicians rely on a multifaceted diagnostic approach, including blood tests, radiomics, comorbidities, tumor markers, and imaging of the primary tumor and lymph nodes, to formulate a comprehensive assessment of tumor progression~\cite{Kim2009Multimodality,Ojiri2020Diagnostic}. Integrating diverse data sources to identify lymph node metastasis (LNM) indicators is challenging due to varying diagnostic methodologies among clinicians, often influenced by personal experience. Therefore, establishing a reliable system for tumor progression staging is essential to ensure consistent and objective clinical decision-making and to customize treatment strategies.

The necessity for utilizing multi-modal data in tumor progression diagnosis stems from its potential to offer a holistic view of the patient's condition. CT imaging serves as a cornerstone for assessing the progression of esophageal cancer, with prior studies demonstrating that models based on CT images of tumors or lymph nodes have achieved favorable outcomes in this task~\cite{jayaprakasam2020role,elsherif2020role,10.1007/978-3-031-16437-8_33,Guo2020A}. Additionally, to delve into deeper details in medical images, some researchers extract radiographic features from specific areas and combine these features with clinical data to enhance the predictive capability for LNM~\cite{huang2016development,wu2017radiomics,wang2019can,sha2020discrimination,cong2020development}. However, these studies have generally simply combined features from different modalities into a new vector~\cite{hu2023multi}, a practice that might neglect the intrinsic medical prior feature relationships between different type of features~\cite{hu2023multi,Feng2019An,Fass2008Imaging} and the complex interactions between data patterns~\cite{Hu2023Enhancing}.

In the actual diagnostic process for cancer metastasis, clinicians often assess the condition of the primary tumor and the proximity of nearby lymph nodes to estimate the current state of metastasis and plan for surgery or radiation therapy~\cite{Marino2019Lymph}. Our Multi-issue Masked Relational Representation Learning (MMRL) strategy specifically addresses this clinical approach by utilizing CT data of tumors with focal lesions and lymph nodes without focal lesions, recognizing the priority of clinical importance between these datasets. Thus, the MMRL strategy is deployed to capture the latent feature relationships between these two types of data, addressing a gap previously overlooked in research.

Following the MMRL strategy, our Heterogeneous Graph Aggregation (HGA) module integrates complex cross-modal relationships via self-attention to enhance multi-modal data fusion. This prepares for our Conditional Feature-guided Diffusion (CFD) process, which refines the diagnostic process by utilizing features aggregated by HGA, including CT images of tumors and lymph nodes, clinical, hematological, and radiomics data to guide the training process. The CFD process simulates the data generation process, progressively reconstructing noise distributions to accurately capture the complete multi-modal conditional distribution of these inputs~\cite{song2020denoising,han2022card,kawar2022denoising,xiao2021tackling}. Also, the randomness inherent in the diffusion process minimizes modality information redundancy~\cite{xiao2021tackling,Veraart2016Denoising}, improving the model's predictive accuracy and reliability~\cite{Hunter1993On}.

The core contributions of our study are as follows: 1) \textbf{We collect a multi-modal dataset of 1,354 cases of ESCC}, utilizing various modalities for lymph node metastasis diagnosis. 2) \textbf{We employ a CFD process based on a multi-modal heterogeneous graph} to explore data relations, effectively reducing redundancy in multi-modal features and leading to optimal model performance. 3) \textbf{We introduce MMRL strategy}, leveraging intra-tissue self-attention and cross-tissue relational mask self-attention for representational learning to explore the priority modeling of inter-tissue relationships and the interactive learning of prognostic information.

\section{Methods}

\subsection{Architecture Overview}
\begin{figure}
\includegraphics[width=\textwidth]{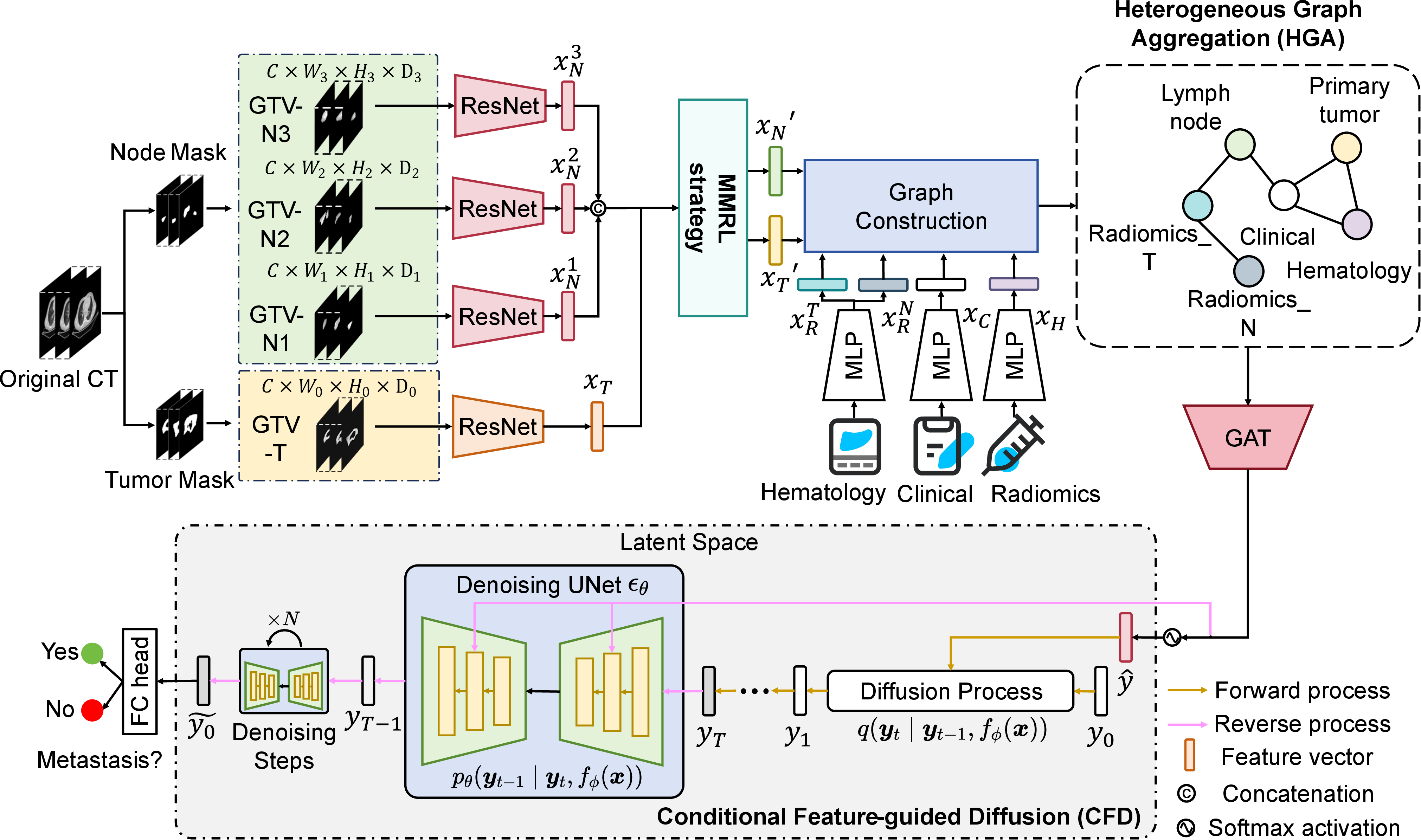}
\caption{Overview framework of our proposed MMFusion. Initially, the MMRL strategy is applied to extract correlation from image representations. Subsequently, we employ HGA to facilitate multi-modal feature interaction. Finally, the CFD method is utilized for feature redundancy elimination.} \label{fig1}
\end{figure}
In this section, the Multi-modal lymph node Metastasis difFusion (MMFusion) model is outlined for LNM diagnosis. Initially, as illustrated in Fig. 1, the gross target volumes (GTV) for the tumor (GTV-T) and lymph node (GTV-N) are derived by applying corresponding masks to the original CT scans. Subsequently, latent imaging representations of GTV-N $\boldsymbol{x}_{N}^{i}, i \in \{1,2,3\}$ and GTV-T $\boldsymbol{x}_{T}$ are extracted using a pretrained ResNet~\cite{chen2019med3d} and processed through the MMRL strategy to discern disease-related information and prognostic correlation priorities between multi-tissue latent representations. These features, along with data from hematology, clinical and radiomics sources, are integrated into a heterogeneous graph. An HGA process, leveraging a Graph Attention Network (GAT)~\cite{velivckovic2017graph}, then identifies potential multimodal feature interactions. Finally, the CFD approach is applied to eliminate multi-modal feature redundancy.

\subsection{Multi-tissue Masked Relational Learning (MMRL) Strategy}

\begin{figure}
\centering
\includegraphics[width=0.8\textwidth]{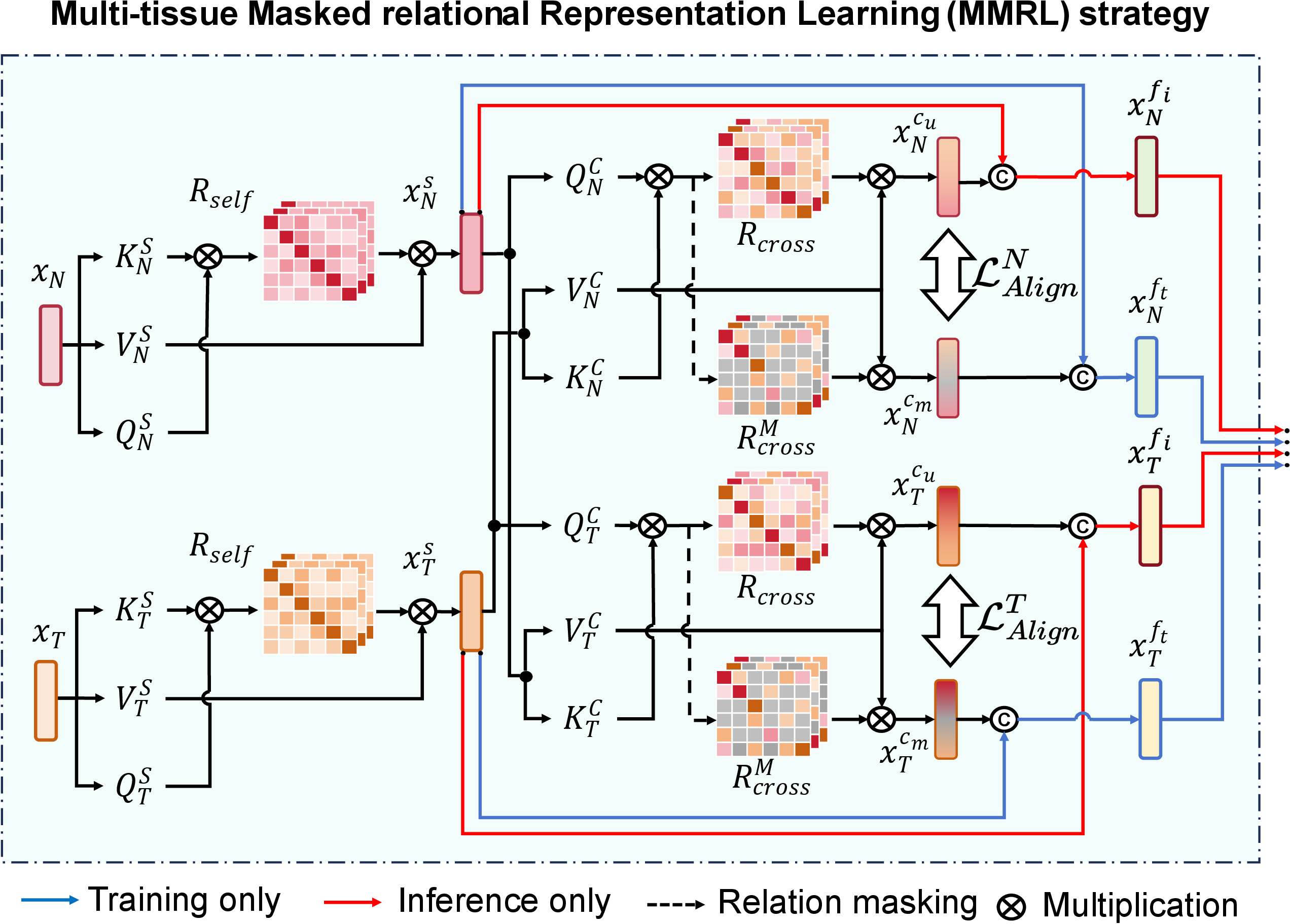}
\caption{Architecture of our proposed Multi-tissue Masked Relational Representation Learning (MMRL) strategy.} \label{fig2}
\end{figure}

The proposed MMRL strategy explores and learns the prognostic-related relationship priorities and interactive information among multi-tissues inspired by~\cite{yang2023mrm}. As shown in Fig.~\ref{fig2}, commencing the exposition with the GTV-N branch as a case study, the methodology initiates with the application of intra-tissue Multi-head Self Attention (MSA)~\cite{vaswani2017attention} to the concatenated GTV-N representation, resulting in the generation of an intra-tissue representation $\boldsymbol{x}_{N}^{S}$. Subsequently, a cross-tissue MSA is conducted. In this phase, the masked relation matrix $\boldsymbol{R}_{cross}^{M}$ is derived by applying random masking to the original relation matrix $\boldsymbol{R}_{cross}$. This process results in the acquisition of both unmasked and masked cross-tissue representations, designated as $\boldsymbol{x}_{N}^{cu}$ and $\boldsymbol{x}_{N}^{cm}$, respectively. For practical application, the masked cross-tissue representation $\boldsymbol{x}_{N}^{cm}$ is merged with $\boldsymbol{x}_{N}^{S}$ during the training phase, whereas $\boldsymbol{x}_{N}^{cu}$ is combined with $\boldsymbol{x}_{N}^{S}$ during the inference phase. To ensure the integrity of feature preservation, an alignment loss is implemented between $\boldsymbol{x}_{N}^{cu}$ and $\boldsymbol{x}_{N}^{cm}$ to enforce a relation modeling constraint.

% consists of relation masking method to mask out cross-tissue feature relations and intact feature preservation to constraint relation modeling. 

\subsection{Multi-modal Heterogeneous graph-based Conditional Feature-guided Diffusion model}
\subsubsection{Heterogeneous Graph Aggregation (HGA) Module}

In graph theory, a heterogeneous graph is generally denoted as $\mathcal{G}=(\mathcal{V}, \mathcal{E}, \mathcal{A}, \mathcal{R})$, with vertex feature matrices $\boldsymbol{F}_{a_i} \in$ $\mathbb{R}^{\left|\mathcal{V}_{a_i}\right| \times d_{a_i}}$ for vertex type $a_i \in \mathcal{A}$, and a task to learn the $d$-dimensional vertex representations $\boldsymbol{h}_v \in \mathbb{R}^d$ for all $v \in \mathcal{V}$ that can capture structural and relational information involved in $\mathcal{G}$. $\mathcal{E}$ $\in$ $\mathcal{R}$ denotes set of edges.
% $\mathcal{E}$ and $\mathcal{R}$ denotes the set of edges and their types.

Our HGA module focuses on aggregating information on other modalities to capture potential interactions between multi-modal features. Thus, we define a heterogeneous graph with only one vertex of each multi-modal data type, $\forall a_i \in \mathcal{A}:\left|\mathcal{V}_{a_i}\right|=1$, with GAT employed to discern the inter-modal interactions. To ensure all possible relationships between different modalities are considered, we utilize a fully connected graph. At each layer, the node feature $\boldsymbol{F}_{a_i}$ will be updated by its attentional neighbor modality to $\boldsymbol{F}_{a_i}^{\prime}$ as follows\cite{zhu2020graph}:
\begin{align}
\boldsymbol{F}_{a_i}^{\prime} = \sigma\left(\sum_{j \in \{i\} \cup \mathcal{N}_i} \boldsymbol{\alpha}_{i j} \boldsymbol{W} \boldsymbol{F}_j\right).
\end{align}

% where $\mathcal{N}_i$ denotes the one-hop neighbors of node $i$, and $\alpha_{i j}$ represents the attention coefficient.

\subsubsection{Conditional Feature-guided Diffusion (CFD) Approach}
For the purpose of eliminating feature redundancy and modeling accurate muti-modality, a novel diffusion model is developed based on CARD~\cite{han2022card}. In the forward process, the Gaussian noise is added to the ground truth $\boldsymbol{y}_{0}$ conditioned by features output from HGA with an arbitrary time step $t \in [1,T]$, which can be defined as:
\begin{align}
    q\left(\boldsymbol{y}_{t} \mid \boldsymbol{y}_{0}, \boldsymbol{f}_{\phi}(\boldsymbol{x})\right)=\mathcal{N}\left(\boldsymbol{y}_{t} ; \sqrt{\bar{\alpha}_{t}} \boldsymbol{y}_{0}+\left(1-\sqrt{\bar{\alpha}_{t}}\right) \boldsymbol{f}_{\phi}(\boldsymbol{x}),\left(1-\bar{\alpha}_{t}\right) \mathbb{I}\right).
\end{align}
where $\bar{\alpha}_{t}=\prod_{t} \alpha_{t}$, $\alpha_{t}=1-\beta_{t}$ with a linear noise schedule $\{\beta_{t}\}_{t=1\dots T} \in (0,1)$. $\mathcal{N}(\cdot, \cdot)$ denotes the Gaussian distribution, $\boldsymbol{f}_{\phi}(\boldsymbol{x})$ denotes the output feature from HGA, and $\mathbb{I}$ is the identity matrix.

In the reverse process, the denoising UNet $\boldsymbol{\epsilon}_{\theta}$~\cite{han2022card} is applied to learn the noise distribution and generate the final prediction $\tilde{\boldsymbol{y}_{0}}$ during denoising steps from a noisy distribution guided by the output features from HGA, which can be defined as:
\begin{align}
    p_{\theta}\left(\boldsymbol{y}_{0: T-1} \mid \boldsymbol{y}_{T}, \boldsymbol{f}_{\phi}(\boldsymbol{x})\right)=\prod_{t=1}^{T} p_{\theta}\left(\boldsymbol{y}_{t-1} \mid \boldsymbol{y}_{t}, \boldsymbol{f}_{\phi}(\boldsymbol{x})\right).
\end{align}
where $\theta$ donates as the parameters in denoising UNet. Concretely, the reverse process in the inference stage can be formalized as:
\begin{align}
     \boldsymbol{y}_{t-1}=\frac{1}{\sqrt{\bar{\alpha}_{t}}}\left(\boldsymbol{y}_{t}-\left(1-\sqrt{\bar{\alpha}_{t}}\right) \boldsymbol{f}_{\phi}(\boldsymbol{x})-\sqrt{1-\bar{\alpha}_{t}} \boldsymbol{\epsilon}_{\theta}\left(\boldsymbol{y}_{t}, \boldsymbol{f}_{\phi}(\boldsymbol{x}), t\right)\right).
\end{align}
where $\bar{\alpha}_{t}$, $\boldsymbol{f}_{\phi}(\boldsymbol{x})$, and $\boldsymbol{\epsilon}_{\theta}$ stand for the same as the previous definition, $\boldsymbol{y}_{t}$ stands for the denoising output from CTD at timestep $t$. By denoising the process under the guidance of the output features $\boldsymbol{f}_{\phi}(\boldsymbol{x})$, the redundancy of multi-modal representations can be effectively eliminated.

\subsection{Optimization Method}
For model optimization, there are two main parts of loss functions. For the non-diffusion part, we utilize Binary Cross-Entropy (BCE) as the primary loss for classification. Also, the Mean Square Error (MSE) loss is utilized for masked and unmasked representational alignment in the MMRL strategy. The loss of the non-diffusion part can be defined as:
\begin{align}
    \mathcal{L}_{\text{non-diff}} = \mathcal{L}_{\text{BCE}}(\hat{\boldsymbol{y}},\boldsymbol{y}_{0}) + \mathcal{L}_{\text{MSE}}^{N}(\boldsymbol{x}_{N}^{c_{u}},\boldsymbol{x}_{N}^{c_{m}}) + \mathcal{L}_{\text{MSE}}^{T}(\boldsymbol{x}_{T}^{c_{u}},\boldsymbol{x}_{T}^{c_{m}}).
\end{align}
where $\hat{\boldsymbol{y}}, \boldsymbol{y}_{0}$ are output predictions from the HGA module and the ground truth. $\boldsymbol{x}_{N}^{c_{u}}, \boldsymbol{x}_{N}^{c_{m}}$ are unmasked and masked intermediate representations of GTV-N in MMRL strategy. For the diffusion part, the denoising UNet $\boldsymbol{\epsilon}_{\theta}$ is utilized to predict the noisy distribution, which is added in the forward process. The loss function of the diffusion part can be defined as:
\begin{align}
    \mathcal{L}_{\text{diff}} = \left\|\boldsymbol{\epsilon}-\boldsymbol{\epsilon}_{\theta}\left( \sqrt{\bar{\alpha}_{t}} \boldsymbol{y}_{0}+\sqrt{1-\bar{\alpha}_{t}} \boldsymbol{\epsilon}+\left(1-\sqrt{\bar{\alpha}_{t}}\right) \boldsymbol{f}_{\phi}(\boldsymbol{x}), \boldsymbol{f}_{\phi}(\boldsymbol{x}), t\right)\right\|^{2}.
\end{align}
where $\boldsymbol{\epsilon}$ stands for Gaussian noise. $\bar{\alpha}_{t}$, $\boldsymbol{f}_{\phi}(\boldsymbol{x})$, $\boldsymbol{y}_{0}$, and $\boldsymbol{\epsilon}_{\theta}$ stand for the same as the previous definition. In total, the final loss function can be defined as:
\begin{align}
    \mathcal{L}_{\text{final}} = \mathcal{L}_{\text{non-diff}} + \mathcal{L}_{\text{diff}}.
\end{align}

% We adopted Pytorch to implement the proposed framework, and the network was trained on NVIDIA RTX 3090 GPU. The batch size was set to $12$, the maximum epoch number was $100$,
% and the initial learning rate was set to $1e^{-4}$. Adam optimizer was utilized and the cosine annealing~\cite{loshchilov2016sgdr} scheduler was applied for learning rate tuning, the minimum learning rate was
% set as $1e^{-5}$ restart from $80$ epoch. The model was trained jointly after warm-up pretraining the non-diffusion part for first $20$ epochs. The timesteps in the diffusion model was set as $T=20$, the noise is linearly scheduled with $\beta_{1} = 1e^{-2}$ and
% $\beta_{T} = 0.95$.

\section{Experiments}   %不要忘了说明我们的图的结构关系^_^ 忘记了。。 -_-|| 凎
\subsection{Dataset Description}
A total of 1,354 ESCC patients are enrolled in this study at Sichuan Cancer Hospital\footnote[1]{This study was approved by the Institutional Ethics Committee of Sichuan Cancer Hospital (No.SCCHEC-02-2020-015).}.
Corresponding data includes preoperative contrast-enhanced CT, clinicopathological parameters, and follow-up information. Contrast-enhanced CT images are retrospectively
acquired from imaging systems manufactured by GE Medical Systems or Philips. To further assess the model's robustness, we execute a three-fold cross-validation. The baseline characteristics of one of the folds can be seen in Table S1.

\subsection{Main Results}

\subsubsection{Comparison Study} We compare our proposed MMFusion against the state-of-the-art (SOTA) LNM diagnostic methods to verify the effectiveness of our method. Models are trained on the training set, selected based on the performance of the validation set, and final results are reported on the test set. As shown in Table
\ref{tab:comparison}, our proposed model achieves the best performance on all metrics compared with other methods. Moreover, a paired t-test is conducted, and the p-value is obtained. Results show that our model significantly outperforms most of the compared methods, which also demonstrates the excellent application prospects of our model in the diagnosis of LNM.

\begin{table*}[!ht]
    \centering
    \caption{Comparison study of the proposed method and other SOTA methods. $\ast$ stands for significant difference (p$<$0.05)}
    \resizebox{\linewidth}{!}{
     \begin{tabular}{c|c|c|c|c}
  \hline\rule{0pt}{10pt}
  {Methods} & {Accuracy (\%)}& \makecell[c]{Precision (\%)} & \makecell[c]{F1 score (\%)}  & \makecell[c]{Recall (\%)} \\
  \hline\rule{0pt}{10pt}
Xie et al.~\cite{xie2022prediction} & 86.35 $\pm$ 1.09$^\ast$ & 76.30 $\pm$ 11.95 & 51.92 $\pm$ 3.25$^\ast$ & 52.83 $\pm$ 1.64$^\ast$  \\
Jin et al.~\cite{jin2021deep} & 87.33 $\pm$ 1.36 & 75.91 $\pm$ 3.24$^\ast$ & 62.84 $\pm$ 2.74$^\ast$ & 55.14 $\pm$ 3.21$^\ast$  \\
 Wan et al.~\cite{wan2023prediction} & 86.22 $\pm$ 0.92$^\ast$ & 71.23 $\pm$ 7.98$^\ast$ & 63.18 $\pm$ 5.87 & 57.98 $\pm$ 7.82 \\
DensePriNet~\cite{zhao2020cross}&86.22 $\pm$ 0.92$^\ast$ & 87.05 $\pm$ 4.12 & 65.21 $\pm$ 3.81 & 54.33 $\pm$ 1.55$^\ast$  \\
Ou et al.~\cite{ou2021ct} & 86.84 $\pm$ 0.17$^\ast$ & 71.85 $\pm$ 4.60$^\ast$ & 61.61 $\pm$ 2.61$^\ast$ & 59.20 $\pm$ 2.02\\
% DeepThyNet~\cite{yao2022deepthy}& & & \\
Huang et al.~\cite{huang2022development} & 87.21 $\pm$ 0.92 & 82.31 $\pm$ 6.31$^\ast$ & 56.51 $\pm$ 0.78$^\ast$ & 55.89 $\pm$ 0.36$^\ast$  \\
\textbf{MMFusion}& \textbf{89.14 $\pm$ 0.99} &  \textbf{93.37 $\pm$ 2.23} & \textbf{71.10 $\pm$ 3.73} & \textbf{65.97 $\pm$ 3.58} \\
  \hline
    \end{tabular} } 
    \label{tab:comparison}
\end{table*}

\subsubsection{Ablation Study} To evaluate the effectiveness of our proposed sub-modules in the MMFusion, we construct three baseline networks. Base1 removes all components from our model, where only vanilla ResNet and MLP are included. Incorporating our MMRL strategy into Base1 yields Base2. Then, by integrating the HGA module into Base2, we obtain Base3. The results, presented in the table~\ref{tab:ablated_}, reveal progressive performance improvements across all metrics with the sequential addition of each proposed module to Base1. This gradual enhancement substantiates the individual contribution of each sub-module to the overarching efficacy of our model.

\begin{table*}[!ht]
    \centering
    \caption{Ablation study of proposed methods. $\ast$ stands for significant difference (p$<$0.05)}
    \resizebox{\linewidth}{!}{
     \begin{tabular}{c|ccc|c|c|c|c}
  \hline\rule{0pt}{10pt}
  {Methods} & {MMRL}& {HGA} & {CFD}  & {\makecell[c]{Accuracy (\%)}}  & {\makecell[c]{Precision (\%)}} & \makecell[c]{F1 score (\%)}  & \makecell[c]{Recall (\%)} \\ 
  \hline\rule{0pt}{10pt}
   Base1& - & - & - & 86.46 $\pm$ 2.11 & 77.59 $\pm$ 1.60$^\ast$  & 64.14 $\pm$ 1.13$^\ast$  & 61.16 $\pm$ 1.16 \\
  Base2& $\checkmark$ & - & - & 87.94 $\pm$ 0.70 & 81.49 $\pm$ 9.16 & 66.90 $\pm$ 4.85 & 65.57 $\pm$ 4.22\\
  Base3& $\checkmark$ & $\checkmark$ & - & 88.65 $\pm$ 0.35 & 84.92 $\pm$ 5.30 & 66.95 $\pm$ 2.66 & 62.46 $\pm$ 2.36 \\
  \textbf{MMFusion} & $\checkmark$ & $\checkmark$ & $\checkmark$ & \textbf{89.14 $\pm$ 0.99} & \textbf{93.37 $\pm$ 2.23} & \textbf{71.10 $\pm$ 3.73} & \textbf{65.97 $\pm$ 3.58} \\
  \hline
    \end{tabular}}
    \label{tab:ablated_}
\end{table*}

\subsubsection{Backbone and Relation Masking Replacement Study} We further verify the performance of our framework under different backbones with and without relation masking. As Table~\ref{tab:backbone} illustrates, integrating relation masking significantly enhances model performance, underscoring the MMRL strategy's ability to capture latent prognostic-related priorities information in multiple tissues.  Table~\ref{tab:backbone} also demonstrates that different backbone networks have little impact on the overall performance. This observation underscores that although backbone architectures lay the groundwork, the actual performance boost comes from the strategic integration and optimization of feature fusion techniques.

\begin{table*}[!ht]
    \centering
    \caption{Backbone and relation masking replacement study of proposed methods.}
     \begin{tabular}{c|c|c|c|c|c|c}
  \hline\rule{0pt}{10pt}
  {CNN} & {GNN}& {MMRL}  & {\makecell[c]{Accuracy (\%)}}  & {\makecell[c]{Precision (\%)}}& \makecell[c]{F1 score (\%)}  & \makecell[c]{Recall (\%)} \\ 
  \hline\rule{0pt}{10pt}
  \multirow{4}*{DenseNet} & \multirow{2}*{GCN} & w/o mask & 88.12 $\pm$ 0.58 & 88.01 $\pm$ 0.95 & 69.52 $\pm$ 6.91 & 63.87 $\pm$ 1.47 \\
  & & w/ mask & 89.01 $\pm$ 0.15 & 88.24 $\pm$ 3.01 & 70.10 $\pm$ 1.01 & 64.21 $\pm$ 0.72 \\
  \cline{2-7}\rule{0pt}{10pt}
  & \multirow{2}*{GAT} & w/o mask & 88.05 $\pm$ 0.52 & 87.53 $\pm$ 3.52 & 66.25 $\pm$ 5.12 & 66.02 $\pm$ 1.05 \\
  & & w/ mask & 88.65 $\pm$ 1.23 & 91.24 $\pm$ 2.11 & 68.15 $\pm$ 1.05 & 65.55 $\pm$ 2.12  \\
\hline\rule{0pt}{10pt}
  \multirow{4}*{\textbf{ResNet}} & \multirow{2}*{GCN} &  w/o mask & 87.15 $\pm$ 1.01 & 90.11 $\pm$ 0.07 & 68.01 $\pm$ 3.70 & 62.71 $\pm$ 2.15  \\
  & &  w/ mask & 87.95 $\pm$ 1.76 & 83.28 $\pm$ 6.84 & 69.12 $\pm$ 3.74 & \textbf{66.50 $\pm$ 3.78} \\
  \cline{2-7}\rule{0pt}{10pt}
  & \multirow{2}*{\textbf{GAT}} &  w/o mask & 88.79 $\pm$ 0.34 & 91.02 $\pm$ 2.77 & 68.11 $\pm$ 5.02 & 63.67 $\pm$ 4.38  \\
  & & \textbf{w/ mask} & \textbf{89.14 $\pm$ 0.99} & \textbf{93.37 $\pm$ 2.23} & \textbf{71.10 $\pm$ 3.73} & 65.97 $\pm$ 3.58 \\
  \hline
    \end{tabular}
    \label{tab:backbone}
\end{table*}

\subsubsection{Visualization of Our Diffusion Procedure}
\begin{figure}
\includegraphics[width=\textwidth]{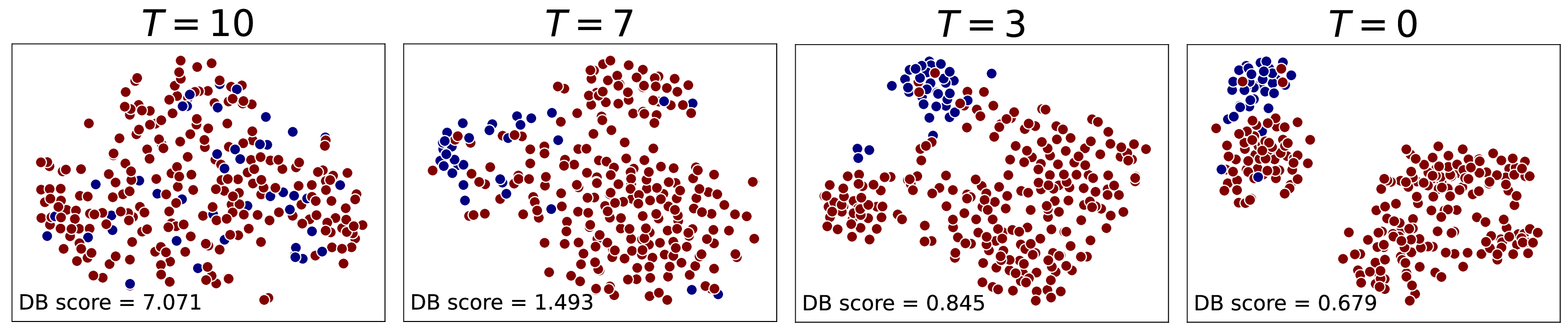}
\caption{Denoised multi-modal feature embedding output from HGA using t-SNE. Red and blue stand for metastasis and non-metastasis, respectively. As the time step encoding process advances, feature redundancy is gradually eliminated, resulting in a clear distribution and a lower DB score, which indicates that our model could effectively perform LNM diagnosis.} \label{fig3}
\end{figure}

To illustrate the reverse process of our CFD, the t-SNE~\cite{van2008visualizing} is applied to visualize the denoised feature embeddings at consecutive time steps. As seen in Fig.~\ref{fig3}, the CFD gradually removes feature redundancy from the multi-modal feature representation output produced by HGA with time step advanced, resulting in a more apparent distribution of classes from the Gaussian distribution. Moreover, we quantify the redundancy elimination effect of the diffusion model using the DB score. As shown in Fig.~\ref{fig3}, the DB score of clustering diminishes over time steps, indicating the progressive elimination of multi-modal feature redundancy and leading to enhanced accuracy in LNM diagnostics.

% lalala.
% (๑•̀ㅂ•́)و✧
% ⬆ 牛逼

\section{Conclusion}
In this work, we introduce MMFusion, a novel framework that integrates primary tumor and lymph node CT images with clinical, hematology, and radiomics data for diagnosing lymph node metastasis in esophageal squamous cell cancer. Specifically, an innovative multi-modal heterogeneous graph-based conditional feature-guided diffusion model is designed to capture the complex interaction within multi-modal representations and to minimize feature redundancy across multi-modalities. Furthermore, we develop a masked relational representation learning strategy to extract prognostic-related information and establish correlation priorities among multiple tissues. Sufficient experiments are conducted and show the effectiveness and superior performance of our proposed method, highlighting its substantial potential to drive advancements in the field.

\begin{credits}
\subsubsection{Acknowledgments} 
This work was supported by National Natural Science Foundation of China (No. 62201323 \& No. 62206242), Natural Science Foundation of Jiangsu Province (No. BK20220266), Zhejiang Provincial Natural Science Foundation of China (No. LDT23F01015F01), National Fund Joint Fund for Regional Innovation and Development (NO. U20A20386), and Science and Technology Department of Sichuan Province (No. 2023YFS0488 \& No. 2023YFQ0055).

\end{credits}


\bibliographystyle{splncs04}
\bibliography{ref}

\begin{thebibliography}{10}
\providecommand{\url}[1]{\texttt{#1}}
\providecommand{\urlprefix}{URL }
\providecommand{\doi}[1]{https://doi.org/#1}

\bibitem{chen2019med3d}
Chen, S., Ma, K., Zheng, Y.: Med3d: Transfer learning for 3d medical image analysis. arXiv preprint arXiv:1904.00625  (2019)

\bibitem{cong2020development}
Cong, M., et~al.: Development of a predictive radiomics model for lymph node metastases in pre-surgical ct-based stage ia non-small cell lung cancer. Lung Cancer  \textbf{139},  73--79 (2020)

\bibitem{elsherif2020role}
Elsherif, S.B., et~al.: Role of precision imaging in esophageal cancer. Journal of Thoracic Disease  \textbf{12}(9), ~5159 (2020)

\bibitem{Fass2008Imaging}
Fass, L.: Imaging and cancer: A review. Molecular Oncology  \textbf{2} (2008). \doi{10.1016/j.molonc.2008.04.001}

\bibitem{Feng2019An}
xia Feng, Q., et~al.: An intelligent clinical decision support system for preoperative prediction of lymph node metastasis in gastric cancer. Journal of the American College of Radiology : JACR  (2019). \doi{10.1016/j.jacr.2018.12.017}

\bibitem{Guo2020A}
Guo, J., et~al.: A prospective analysis of the diagnostic accuracy of 3t mri, ct and endoscopic ultrasound for preoperative t staging of potentially resectable esophageal cancer. Cancer Imaging  \textbf{20} (2020). \doi{10.1186/s40644-020-00343-w}

\bibitem{han2022card}
Han, X., Zheng, H., Zhou, M.: Card: Classification and regression diffusion models. Advances in Neural Information Processing Systems  \textbf{35},  18100--18115 (2022)

\bibitem{hu2023multi}
Hu, D., Li, S., Wu, N., Lu, X.: A multi-modal heterogeneous graph forest to predict lymph node metastasis of non-small cell lung cancer. IEEE Journal of Biomedical and Health Informatics  \textbf{27}(3),  1216--1224 (2023)

\bibitem{Hu2023Enhancing}
Hu, Z., et~al.: Enhancing the accuracy of lymph-node-metastasis prediction in gynecologic malignancies using multimodal federated learning: Integrating ct, mri, and pet/ct. Cancers  \textbf{15} (2023). \doi{10.3390/cancers15215281}

\bibitem{huang2022development}
Huang, C., et~al.: Development and validation of a deep learning model to predict survival of patients with esophageal cancer. Frontiers in Oncology  \textbf{12},  971190 (2022)

\bibitem{huang2016development}
Huang, Y.q., et~al.: Development and validation of a radiomics nomogram for preoperative prediction of lymph node metastasis in colorectal cancer. Journal of clinical oncology  \textbf{34}(18),  2157--2164 (2016)

\bibitem{Hunter1993On}
Hunter, J., Craig, P., Phillips, H.: On the use of random walk models with spatially variable diffusivity. Journal of Computational Physics  \textbf{106},  366--376 (1993). \doi{10.1016/S0021-9991(83)71114-9}

\bibitem{jayaprakasam2020role}
Jayaprakasam, V.S., et~al.: Role of imaging in esophageal cancer management in 2020: update for radiologists. American Journal of Roentgenology  \textbf{215}(5),  1072--1084 (2020)

\bibitem{jin2021deep}
Jin, C., et~al.: Deep learning analysis of the primary tumour and the prediction of lymph node metastases in gastric cancer. British Journal of Surgery  \textbf{108}(5),  542--549 (2021)

\bibitem{kawar2022denoising}
Kawar, B., Elad, M., Ermon, S., Song, J.: Denoising diffusion restoration models. Advances in Neural Information Processing Systems  \textbf{35},  23593--23606 (2022)

\bibitem{Kim2009Multimodality}
Kim, T., Kim, H.Y., Lee, K., Kim, M.S.: Multimodality assessment of esophageal cancer: preoperative staging and monitoring of response to therapy. Radiographics : a review publication of the Radiological Society of North America, Inc  \textbf{29 2},  403--21 (2009). \doi{10.1148/rg.292085106}

\bibitem{van2008visualizing}
Van~der Maaten, L., Hinton, G.: Visualizing data using t-sne. Journal of machine learning research  \textbf{9}(11) (2008)

\bibitem{Marino2019Lymph}
Marino, M., Avendaño, D., Zapata, P., Riedl, C., Pinker, K.: Lymph node imaging in patients with primary breast cancer: Concurrent diagnostic tools. The Oncologist  \textbf{25},  e231 -- e242 (2019). \doi{10.1634/theoncologist.2019-0427}

\bibitem{Ojiri2020Diagnostic}
Ojiri, H.: Diagnostic imaging of the esophageal cancer. Esophageal Squamous Cell Carcinoma  (2020). \doi{10.1007/978-4-431-54977-2_3}

\bibitem{ou2021ct}
Ou, J., et~al.: Ct radiomics features to predict lymph node metastasis in advanced esophageal squamous cell carcinoma and to discriminate between regional and non-regional lymph node metastasis: a case control study. Quantitative Imaging in Medicine and Surgery  \textbf{11}(2), ~628 (2021)

\bibitem{sha2020discrimination}
Sha, X., Gong, G., Qiu, Q., Duan, J., Li, D., Yin, Y.: Discrimination of mediastinal metastatic lymph nodes in nsclc based on radiomic features in different phases of ct imaging. BMC medical imaging  \textbf{20}(1), ~1--8 (2020)

\bibitem{song2020denoising}
Song, J., Meng, C., Ermon, S.: Denoising diffusion implicit models. arXiv preprint arXiv:2010.02502  (2020)

\bibitem{sung2021global}
Sung, H., et~al.: Global cancer statistics 2020: Globocan estimates of incidence and mortality worldwide for 36 cancers in 185 countries. CA: a cancer journal for clinicians  \textbf{71}(3),  209--249 (2021)

\bibitem{vaswani2017attention}
Vaswani, A., et~al.: Attention is all you need. Advances in neural information processing systems  \textbf{30} (2017)

\bibitem{velivckovic2017graph}
Veli{\v{c}}kovi{\'c}, P., et~al.: Graph attention networks. arXiv preprint arXiv:1710.10903  (2017)

\bibitem{Veraart2016Denoising}
Veraart, J., Novikov, D., Christiaens, D., Ades-aron, B., Sijbers, J., Fieremans, E.: Denoising of diffusion mri using random matrix theory. NeuroImage  \textbf{142},  394--406 (2016). \doi{10.1016/j.neuroimage.2016.08.016}

\bibitem{wan2023prediction}
Wan, L., et~al.: Prediction of lymph node metastasis in stage t1--2 rectal cancers with mri-based deep learning. European Radiology  \textbf{33}(5),  3638--3646 (2023)

\bibitem{wang2019can}
Wang, X., et~al.: Can peritumoral radiomics increase the efficiency of the prediction for lymph node metastasis in clinical stage t1 lung adenocarcinoma on ct? European radiology  \textbf{29},  6049--6058 (2019)

\bibitem{wu2017radiomics}
Wu, S., et~al.: A radiomics nomogram for the preoperative prediction of lymph node metastasis in bladder cancer. Clinical Cancer Research  \textbf{23}(22),  6904--6911 (2017)

\bibitem{xiao2021tackling}
Xiao, Z., Kreis, K., Vahdat, A.: Tackling the generative learning trilemma with denoising diffusion gans. arXiv preprint arXiv:2112.07804  (2021)

\bibitem{xie2022prediction}
Xie, C., et~al.: Prediction of individual lymph node metastatic status in esophageal squamous cell carcinoma using routine computed tomography imaging: comparison of size-based measurements and radiomics-based models. Annals of Surgical Oncology  \textbf{29}(13),  8117--8126 (2022)

\bibitem{yang2023mrm}
Yang, Q., Li, W., Li, B., Yuan, Y.: Mrm: Masked relation modeling for medical image pre-training with genetics. In: Proceedings of the IEEE/CVF International Conference on Computer Vision. pp. 21452--21462 (2023)

\bibitem{10.1007/978-3-031-16437-8_33}
Yao, J., Ye, X., Xia, Y., Zhou, J., Shi, Y., Yan, K., Wang, F., Lin, L., Yu, H., Hua, X.S., Lu, L., Jin, D., Zhang, L.: Effective opportunistic esophageal cancer screening using noncontrast ct imaging. In: Wang, L., Dou, Q., Fletcher, P.T., Speidel, S., Li, S. (eds.) Medical Image Computing and Computer Assisted Intervention -- MICCAI 2022. pp. 344--354. Springer Nature Switzerland, Cham (2022)

\bibitem{zhao2020cross}
Zhao, X., et~al.: A cross-modal 3d deep learning for accurate lymph node metastasis prediction in clinical stage t1 lung adenocarcinoma. Lung Cancer  \textbf{145},  10--17 (2020)

\bibitem{zhu2020graph}
Zhu, S., Pan, S., Zhou, C., Wu, J., Cao, Y., Wang, B.: Graph geometry interaction learning. Advances in Neural Information Processing Systems  \textbf{33},  7548--7558 (2020)

\end{thebibliography}

\end{document}

\title{Supplementary material: MMFusion: Multi-modality Diffusion Model for Lymph Node Metastasis Diagnosis in Esophageal Cancer}
\author{}
\institute{}
\maketitle
% \vspace{-15mm}
\section{Basic information and division of dataset}
% \vspace{-6mm}

\begin{table}
\centering
    \caption{Baseline characteristics of the study cohorts}
     \begin{tabular}{lccc}
  \toprule
  \multirow{2}*{Characteristic} & \makecell[c]{Training cohort} & \makecell[c]{Validation cohort} & \makecell[c]{Test cohort}   \\
  & (n=947) & (n=136) & (n=271) \\ 
  \midrule
  \textbf{OStime} (\%)  \\
   $\ge$ 35 month & 443 (46.8) & 65 (47.7) & 112 (41.3)\\
   $<$ 35 month & 504 (53.2) & 71 (52.2) &  149 (54.9)\\ 
  % \textbf{OSstatue} (\%) \\ 
  % \hspace{2em} censoring & 550 (50.8) & 142 (52.4)\\ \hspace{2em} event & 533 (49.2) &  129 (47.6)\\
  \textbf{Age} (\%) \\
   $\ge$ 70 & 560 (59.1) & 75 (55.1)& 166 (61.2)\\ 
   $<$ 70 & 387 (40.8) & 61 (44.8) & 105 (38.7)\\
  \textbf{Sex} (\%) \\ 
  male & 791 (83.5) & 112 (82.3) & 227 (83.8)\\ 
  female & 156 (16.4) & 24 (17.6) &  44 (16.2)\\ 
  \textbf{KPS} (\%)\\ 
  G1 & 1 (0.1) & 1 (0.7) & 0 (0.0)\\ 
   G2 & 528 (55.7) & 66 (48.5) & 145 (53.5) \\ 
   G3 &  414 (43.7) & 68 (50.0) & 124 (45.8)\\ 
 G4 & 4 (0.4) & 1 (0.7) & 2 (0.7) \\ 

 \bottomrule
    \end{tabular}
    \label{tab:clinicalbaseline}
\end{table}

\newpage
\section{Training and implementation details}
\begin{table}
\centering
    \caption{Training and implementation details}
     \begin{tabular}{lll}
  \toprule
 Type & Parameters & Value    \\ 
  \midrule
  \multirow{15}*{Overall}& GPU & NVIDIA RTX 3090 \\
 & Batch size & 12 \\
 & Optimizer & Adam \\
 & Weight decay & $5e-4$ \\
 & Learning rate & $1e-4$ \\
 & Epochs & 100\\
 & Scheduler & Cosine Annealing \\
 & Restart epoch & 80 \\
 & Min learning rate & $1e-5$ \\
 & Augmentation & RandomNoise (p=0.4)\\
 & & RandomBiasField (p=0.3) \\
 & & RandomFlip (p=0.6) \\
 & & RandomMotion (p=0.2) \\
 & Layers of GAT & 1\\
 & Backone of CNN & ResNet50 \\
 & \makecell[l]{Guidance model warm-up \\ training epochs} & 50 \\
 & Masking ratio & 15$\%$ \\
 \midrule
 \multirow{3}*{CTD} & timesteps & 10 \\
  & $\beta_{1}$ & $0.01$ \\
  & $\beta_{T}$ & $0.95$ \\
 % & 
 \bottomrule
    \end{tabular}
    \label{tab}
\end{table}